\documentclass[fleqn,twoside]{article}
\usepackage{espcrc2}

\usepackage{graphicx}
\usepackage[figuresright]{rotating}

\newcommand{\AmS}{{\protect\the\textfont2
  A\kern-.1667em\lower.5ex\hbox{M}\kern-.125emS}}

\title{On a possible photon origin of the most-energetic AGASA events}

\author{P. Homola\address[IFJ]{Institute of Nuclear Physics PAN, ul. Radzikowskiego 152,
    31-342 Krak\'ow, Poland},
        M. Risse\addressmark[IFJ]$^,$\address[FZK]{Forschungszentrum Karlsruhe, Institut fuer Kernphysik,
    76021 Karlsruhe, Germany},
        R. Engel\addressmark[FZK],
        D. G\'ora\addressmark[IFJ],
	D. Heck\addressmark[FZK],
	J. P\c{e}kala\addressmark[IFJ], 
	B. Wilczy\'nska\addressmark[IFJ]
        and 
	H.Wilczy\'nski\addressmark[IFJ]}

\begin{document}

\begin{abstract}
In this work the ultra high energy cosmic ray events recorded by the AGASA experiment 
are analysed. With detailed simulations of the extensive air showers
initiated by photons, the probabilities are determined of the photonic origin of the 6 
AGASA events for which the muon densities were measured and the reconstructed energies 
exceeded 10$^{20}$~eV.
On this basis a new, preliminary upper limit on the photon fraction in 
cosmic rays above 10$^{20}$~eV is derived and compared to 
the predictions of exemplary top-down cosmic-ray origin models.

\vspace{1pc}
\end{abstract}

\maketitle

\section{Photons as UHECR?}
Studies on ultra-high energy photons as primary cosmic rays have the potential
to discern between different models of origin of ultra-high energy cosmic rays (UHECR). 
Many top-down scenarios predict a dominance of photons in the CR flux above 10$^{20}$~eV (for a review
see~\cite{topdown}), while according to classical acceleration scenarios in this energy range
the CR flux should consist mainly of hadrons. Thus, any experimental conclusions about a
fraction of photons in UHECR, e.g. an upper limit on this fraction, are expected
to provide constraints for the models of UHECR origin. 

Presently, there are few experimental upper limits on photon fraction estimated on the 
basis of cosmic rays of energies above 10$^{19}$~eV (\cite{shin02}, \cite{haverah-photon-limit}). 
They are plotted in Fig.~\ref{plot-shin02-ul-ph} and compared to the predictions of the three exemplary
top-down models: super heavy dark matter (SHDM)~\cite{shdm-mr15}, Z-bursts (ZB) and topological defects 
(TD)~\cite{mr33}. It can be seen that the available limits are not very constraining
since below 5$\times$10$^{19}$~eV a typical predicted photon fraction is still low. 
Above 10$^{20}$~eV, where the photon
fraction predictions are larger, there is no experimental upper limit, although
there are a few well documented cosmic ray events of such extremely large energies.  
While large amount of UHECR data from next generation giant cosmic-ray detectors 
is still on their way, in this work the available data of extensive air showers detected 
by the AGASA experiment are analysed to derive an upper limit on the photon fraction in UHECR. 
In order to determine the probability of an event being initiated by a photon
the muon densities measured on ground, which are expected to be good 
indicators of primary mass~\cite{bert,plya}, were compared to the simulations.  

\section{AGASA cosmic ray data above 10$^{20}$ eV}
\label{sec-results-agasa}

Among the 11 cosmic ray events above 10$^{20}$~eV recorded by the 
AGASA experiment~\cite{agasa-shin02,shin02} in 6 cases
the muon densities at a distance of 1000 m from the shower core ($\rho_\mu(1000)$) were
measured and published in Ref.~\cite{shin02} (see Table~\ref{tab-agasa-sim}).
The accuracy of $\rho_\mu(1000)$ was evaluated to be $\sim40\%$~\cite{shin02}.

\subsection{Simulations} 
The published energies and arrival directions were used for the simulation of 
photon-induced air showers. The effect of pre-cascading of photons in the geomagnetic 
field (preshower effect)~\cite{erber,mcbr} and the subsequent air shower simulations
were performed with use of the CORSIKA package~\cite{corsika618} 
with coupled PRESHOWER code~\cite{preshcors}.
For each of the 6 events with available experimental $\rho_\mu(1000)$ values,
100 photon-induced showers were simulated. This allowed
obtaining 6 distributions of simulated muon densities $\rho_{\mu,sim}(1000)$
at the observational level.

The exact values of the reconstructed primary energies $E_0$ were published
in Ref. \cite{agasa-evlist} and updated on the AGASA web 
page~\cite{agasa-www}. These values were obtained with accuracy of 25\%, assuming that
all the showers were initiated by hadrons. As noted in Ref.~\cite{shin02}, 
for photon-induced showers, $E_0$ underestimates the real primary energy by about 20\%.
This 20\% correction was applied in all the photon-induced shower simulations presented here. 

\vspace{-0.1cm}
\subsection{Probabilities of photonic origin}
The basic approach to determine the probability of a primary photon origin of an individual event 
is quite similar to the one outlined in Ref.~\cite{mr-fe}.
The simulated distributions of $\rho_{\mu,sim}(1000)$ were compared to the experimental values
in the following way.
It is assumed that the uncertainties connected to the 
measurements of the muon density $\rho_\mu(1000)$ for each AGASA
event follow the Gaussian distribution
with the uncertainty $\sigma_{\rho_\mu(1000)}=0.4\rho_\mu(1000)$.
The simulations for the energy and arrival direction of the j-th recorded event 
gave $n=100$ expected values $\rho_{\mu,sim}^{ij}(1000)$, where index $i$ denotes
the i-th simulated shower.
The probability $P^{ij}$ that $\rho_\mu^j(1000)$ differs by more than $\delta^{ij}$ from the 
expected muon density $\rho_{\mu,sim}^{ij}(1000)$ is given by 
\begin{eqnarray}
\label{eq-nmu-prob}
P^{ij} \equiv P(|\rho_{\mu,sim}^{ij}(1000)-\rho_\mu^j(1000)|>\delta^{ij})=
\nonumber \\
=1-\int^{\rho_\mu^{j}(1000)+\delta^{ij}}_{\rho_\mu^{j}(1000)-\delta^{ij}}f(x,\rho_\mu^{j}(1000))dx,
\end{eqnarray}
with $f(x,\rho_\mu^{j}(1000))$ being the Gaussian distribution. 
Since $\sigma_{\rho_\mu} \approx 0.4\rho_\mu$, $f$ 
has a small (less than 1\% of $\int f(x)dx$) contribution from negative values of muon densities $x$.
This contribution was accounted for by a simple renormalization of $P^{ij}$.
The probability for the j-th event being initiated by 
an UHE photon, $P_\gamma^j$, was obtained by taking the average $P^{ij}$ values.
The experimental data together with the results of the simulations and analysis are collected
in Table \ref{tab-agasa-sim}.
\begin{table}
\begin{center}
\caption{Average muon densities $\langle \rho_{\mu,sim}^{ij}(1000) \rangle$ 
and their RMS fluctuations simulated at ground level for the AGASA 
events with $E_0 > 10^{20}$~eV are compared to the experimental values $\rho_\mu^j(1000)$. 
The $E_0$ values are increased by 20\% with respect to the published data.
The probabilities $P_\gamma^j$ that the j-th event was initiated by an UHE photon are given.
The azimuths of the arrival directions increase clockwise from North.
For the analysis method see the text.}
\vspace{0.2cm}
\begin{tabular}
{p{2.5cm}p{0.35cm}p{0.35cm}p{0.35cm}p{0.35cm}p{0.35cm}p{0.35cm}} \hline
\multicolumn{7}{c}{AGASA data}\\
\hline
$E_0$ [EeV] & 295 & 240 & 173 & 161 & 126 & 125 \\
zenith [$^\circ$] & 37 & 23 & 14 & 35 & 33 & 37 \\
azimuth [$^\circ$] & 260 & 236 & 211 & 55 & 108 & 279 \\ 
$\rho_\mu^j(1000)$ [m$^{-2}$] & 8.9 & 10.7 & 8.7 & 5.9 & 12.6 & 9.3 \\ \hline
\multicolumn{7}{c}{simulations}\\
\hline
preshowers [\%] & 100 & 100 & 96 & 100 & 93 & 100 \\
$\rho_{\mu}^{j,s}(1000)$ [m$^{-2}$] & 4.3 & 3.1 & 2.1 & 2.3 & 1.7 & 1.8 \\
RMS($\rho_{\mu}^{j,s}$)[m$^{-2}$] & 0.7 & 0.6 & 0.8 & 0.3 & 0.4 & 0.3 \\
$P_\gamma^j$ [\%] & 19.5 & 7.3 & 6.0 & 12.3 & 2.4 & 3.8 \\
\hline 
\end{tabular}\\
\label{tab-agasa-sim}
\end{center}
\vspace{-1.0cm}
\end{table}
For each event, the probabilities $P_\gamma^j$ are small, but not negligible. 

\section{Upper limit on photon fraction in UHECR above 10$^{20}$~EeV}

After the probabilities $P_\gamma^j$ have been determined, the upper limit for the photon fraction in
the cosmic rays with $E_0 > 100$ EeV can be estimated in the following way.
Let us calculate the probability that $N_\gamma$ out of 6 AGASA events were induced by photons:
\begin{equation}
\label{p-gamma-frac}
P(N_\gamma=n)=\sum \prod_{k=1}^n P_\gamma^{j_k}\prod_{l=1}^{6-n}(1-P_\gamma^{i_l})
\end{equation}
where $j_k,i_l \in \mathbf{A}=\{1,...,6\}$, $\forall~k,l \in \mathbf{A}:j_k \ne j_l$  
and the summation extends over all $n$-elemental combinations of indices $j_k$ in $\mathbf{A}$.
The values of $P(N_\gamma=n)$ are collected in Table 
\ref{tab-gamma-frac}.
\begin{table}
\begin{center}
\caption{Probability that $N_\gamma$ out of the 6 AGASA events 
were initiated by a photon.}
\begin{tabular}
{p{0.7cm}p{0.45cm}p{0.45cm}p{0.45cm}p{0.45cm}p{0.55cm}p{0.55cm}p{0.55cm}} \hline
$N_\gamma$ & ~~0 & ~~1 & ~~2 & ~~3 & ~~4 & ~~5 & ~~6  \\ \hline 
$P$[\%] &  57.8 & 34.0 & 7.4 & 0.8 & \hspace{-0.2cm}4E-2 & \hspace{-0.2cm}1E-3 & \hspace{-0.2cm}1E-5     \\
\hline
\end{tabular}\\
\label{tab-gamma-frac}
\end{center}
\vspace{-1.0cm}
\end{table}
These values determine the upper limit on gamma fraction in cosmic rays.   
The probability that less than 3 out of the analysed 6 AGASA events with $E_0 > 100$~EeV 
were initiated by photons is larger than 95\%, i.e. $P(N_\gamma/N < 3/6) > 95\%$.
This allows the conclusion that the upper limit on the photon fraction in cosmic rays above $10^{20}$ eV 
equals $3/6 = 50\%$ at 95\% confidence level. 

\section{Discussion}

\begin{figure}
\vspace{-0.3cm}
\begin{center}
\includegraphics[width=0.4\textwidth,angle=0]{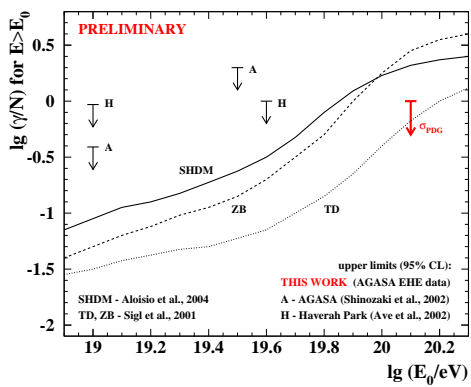}
\end{center}
\vspace{-1.3cm}
\caption {The ratio of photon-induced showers to 
hadron-induced ones as a function of the primary particle energy~\cite{shin02}. The upper limits
obtained by the AGASA and Haverah Park experiments at 95\% CL are indicated by arrows. For explanation
of the theoretical models see the text.}
\vspace{-0.6cm}
\label{plot-shin02-ul-ph}
\end{figure}

The obtained 50\% upper limit on gamma fraction corresponds to $N_\gamma/N_{nucl}(>10^{20}$eV$) < 1$.
In Fig.~\ref{plot-shin02-ul-ph}, the above result is compared to the theoretical 
expectations of exemplary top-down models of cosmic ray origin and to the previous experimental
upper limits on the photon fraction in UHECR of energies above 10$^{20}$~eV.  
The models that predict gamma fractions higher than the obtained upper limit
are disfavored by the AGASA data.

It should be noted that a considerable systematic uncertainty is related
to extrapolating the photonuclear cross-section $\sigma_{\gamma-air}$ to highest energies~\cite{mr-pylos}. 
In relation with the current 
analysis, assuming that $\sigma_{\gamma-air}$ increases more rapidly at highest energies would result
in larger muon densities on ground and hence in larger
probabilities of photonic origin of the analysed events. This will be discussed elsewhere.

The presented analysis method of determining the upper limit of photon fraction can be used as well
for a larger data sample which is expected soon from the new
generation cosmic ray detectors. The conclusions based on these new data should offer
a good way to distinguish between different models of cosmic ray origin. 

{\it Acknowledgments.} The authors thank K.~Shinozaki for helpful discussions.
This work was partially supported by the Polish State Committee for
Scientific Research under grants No.~PBZ~KBN~054/P03/2001 and 2P03B~11024 and in Germany by the DAAD
under grant No.~PPP~323. 
MR is supported by the Alexander von Humboldt-Stiftung.

\end{document}